\documentclass[preprint,12pt]{elsarticle}

\usepackage{amssymb,hyperref}

\journal{Nuclear Instruments and Methods}

\begin{document}

\begin{frontmatter}

\title{Design and construction of the POLAR detector}
\author[isdc]{N.~PRODUIT\corref{cor1}}
\ead{Nicolas.Produit@unige.ch}
\author[ihep]{T.W.~BAO}
\author[pol]{T.~BATSCH}
\author[isdc]{T.~BERNASCONI}
\author[psi]{I.~BRITVICH}
\author[dpnc]{F.~CADOUX}
\author[isdc]{I.~CERNUDA}
\author[ihep]{J.Y.~CHAI}
\author[ihep]{Y.W.~DONG}
\author[isdc]{N.~GAUVIN}
\author[psi]{W.~HAJDAS}
\author[dpnc]{M.~KOLE}
\author[ihep]{M.N.~KONG}
\author[psi]{R.~KRAMERT}
\author[ihep]{L.~LI}
\author[ihep]{J.T.~LIU}
\author[ihep]{X.~LIU}
\author[psi]{R.~MARCINKOWSKI}
\author[dpnc]{S.~ORSI}
\author[dpnc]{M.~POHL}
\author[dpnc]{D.~RAPIN}
\author[pol]{D.~RYBKA}
\author[psi]{A.~RUTCZYNSKA}
\author[ihep]{H.L.~SHI}
\author[psi]{P.~SOCHA}
\author[ihep]{J.C.~SUN}
\author[ihep]{L.M.~SONG}
\author[pol]{J.~SZABELSKI}
\author[psi]{I.~TRASEIRA}
\author[psi]{H.L.~XIAO}
\author[ihep]{R.J.~WANG}
\author[ihep]{X.~WEN}
\author[ihep]{B.B.~WU}
\author[ihep]{L.~ZHANG}
\author[ihep]{L.Y.~ZHANG}
\author[ihep]{S.N.~ZHANG}
\author[ihep]{Y.J.~ZHANG}
\author[pol]{A.~ZWOLINSKA}

\address[isdc]{University of Geneva, Geneva Observatory, ISDC,
16, Chemin d'Ecogia, 1290 Versoix Switzerland}
\address[dpnc]{University of Geneva, D\'epartement de physique Nucl\'eaire et Corpusculaire (DPNC), quai Ernest-Ansermet 24, 1205 Geneva, Switzerland}
\address[psi]{Paul Scherrer Institut
5232 Villigen PSI, Switzerland}
\address[ihep]{Key Laboratory of Particle Astrophysics, Institute of High Energy Physics, Chinese Academy of Sciences, Beijing, China, 100049}
\address[pol]{National Centre for Nuclear Research
ul. A. Soltana 7, 05-400 Otwock, Swierk, Poland}

\cortext[cor1]{Corresponding author}

\begin{abstract}
The POLAR detector is a space based Gamma Ray Burst (GRB) polarimeter
with a wide field of view, which covers almost half the sky. The instrument uses Compton scattering of
gamma rays on a plastic scintillator hodoscope to measure the polarization of the incoming photons. The instrument has been successfully launched on board of the Chinese space laboratory Tiangong~2 on September 15, 2016. The construction of the instrument components is described in this article. Details are provided on problems encountered during the construction phase and their solutions.  Initial performance of the instrument in orbit is as expected from ground tests and Monte Carlo simulation. 
\end{abstract}

\begin{keyword}
  Polarimeter \sep X-rays \sep Compton scattering
\end{keyword}
\end{frontmatter}
\section{Introduction}
POLAR is a GRB polarimeter, which implements the conceptual design described in reference
\citep{Nicolas}. 
The basic concept is to  measure the mean degree of polarization and the azimuthal angle of the polarization vector by analyzing the angular distribution of the azimuthal Compton scattering angle of a sample of photons associated with a GRB. This requires that both the recoil electron from the Compton process and a secondary interaction of the outgoing photon are observed. POLAR uses the same plastic scintillator material for both, and determines the azimuthal scattering angle by structuring the target into 40$\times$40 pixels read out separately.  The detector is sensitive to gamma rays with an energy between a 10 keV and 500 keV. Its field of view covers almost half the sky.  

POLAR was chosen by the Chinese authorities to be one of the instruments on board of the Chinese space laboratory Tiangong~2 (TG-2). The laboratory and its payloads were successfully launched on September 15, 2016 into a low Earth orbit (300 km altitude 45 degree inclination).

\section{POLAR models}
Following this design concept, a demonstrator with only 8$\times$8 pixels was first produced and tested in a polarized gamma ray beam at the European Synchrotron Radiation Facility (ESRF) as described in \citep{Orsi}. These tests demonstrated a good agreement between the produced hardware performance and the Monte-Carlo results. 

Several models of the complete POLAR detector were then produced. A qualification model (QM), was used to demonstrate the performance of the instrument at a beam test at ESRF \citep{Orsi}. Afterwards this QM underwent all the qualification procedures performed at SERMS Terni.
This instrument qualification consisted of both random and sinusoidal vibration tests, electromagnetic compatibility (EMC) tests, a thermal cycle test, a thermal vacuum test, and shock tests in the direction perpendicular to the bars. During the last shock test, consisting of a 500 $g$ shock in the direction
along the scintillation bars, the instrument suffered a catastrophic mechanical failure. 

After this incident, the shock mitigation concept was fully redesigned as described in Section~\ref{sec:design}.
All the individual design changes where separately tested using dedicated vibration and shock equipment. The upgraded model underwent all qualification tests successfully, including a shock test in the Z direction. This model was subsequently used as the flight spare model (FMS).

A copy of this instrument was produced as the flight model (FM). A batch of less stringent flight acceptance tests were performed on this model. The FM was fully calibrated using a 100\% polarized beam at the ESRF facility. The calibration of the flight model will be described elsewhere \citep{Yongjie}. 

Using the GEANT4 package \citep{GEANT}, a complete model of the POLAR set-up was constructed to provide a high reliable simulation of its properties and performance. The fidelity of the model and residual systematic were demonstrated using data from the ESRF calibration of the FM \citep{MerlinSim}. 

\section{Overall design}\label{sec:design}
As GRBs appear randomly in the sky and for a very short instant (milliseconds to minutes), a large field of view (FoV) is required for the polarimeter
in order to detect as many GRBs as possible. The FoV of POLAR is almost half of the sky, as detailed in \citep{Nicolas}.

POLAR is mounted on the Tiangong~2 Chinese space laboratory which was launched into Low Earth Orbit on September 15, 2016.
The instrument is composed of two parts, the OBOX assembly, which consists of the target and associated electronics and the IBOX assembly which contains ancillary instrumentation as well as the space craft interfaces. The two assemblies are connected by a harness for power, data and command transmission. 

\section{OBOX assembly}
\begin{figure}
  \centering
  \includegraphics[width=\linewidth]{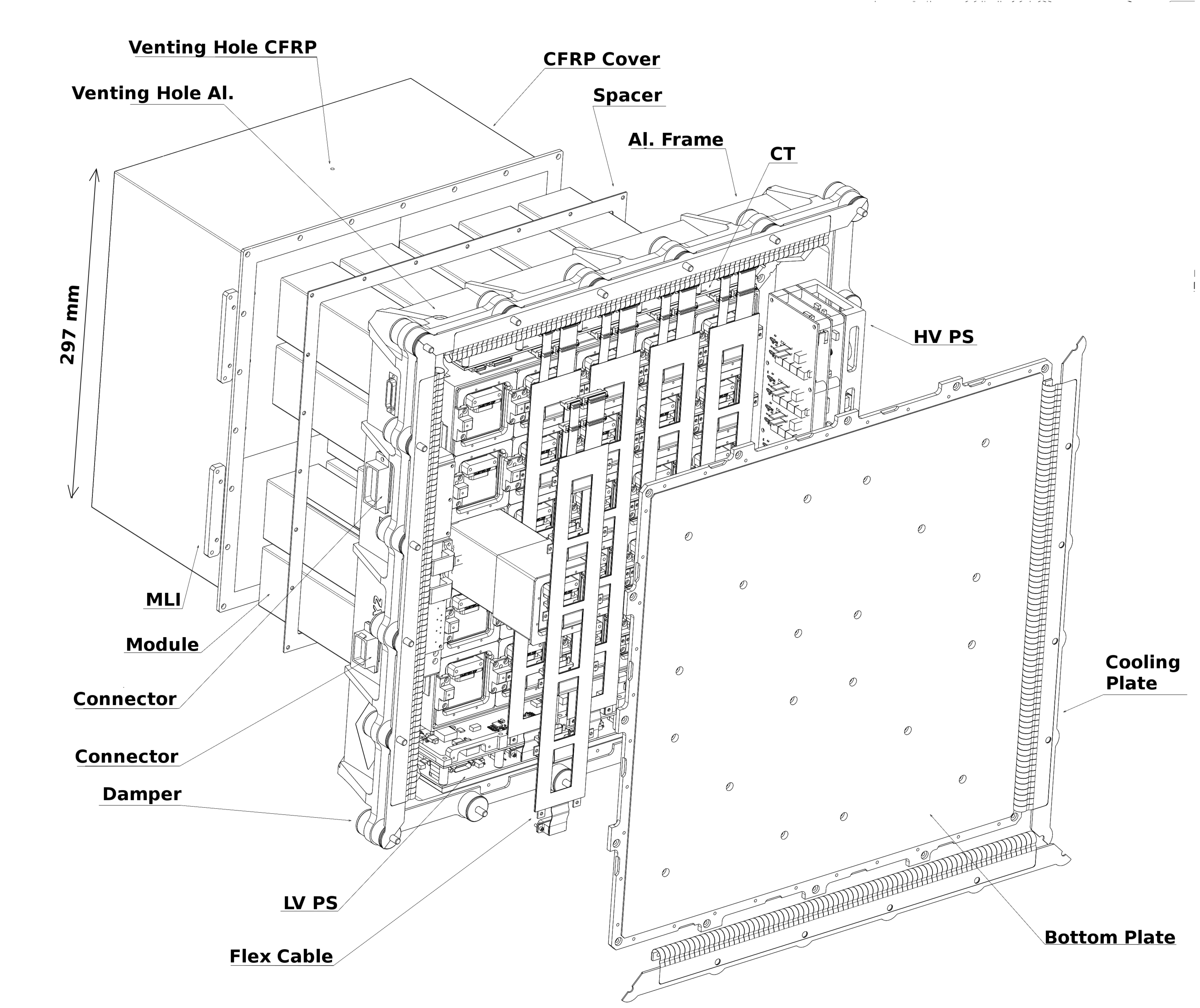}
  \caption{OBOX assembly}
  \label{fig:obox}
\end{figure}

OBOX is mounted on the outer surface of TG-2 through 20 shock absorbers as shown in figure \ref{fig:obox}. It is connected to IBOX with two electrical cable bundles for separate power and data transmission. OBOX is composed of the following parts displayed in the figure and detailed in the following: 
\begin{itemize}
\item the aluminum frame containing the electronics (POLAR GRID);
\item the target with its 25 modules;
\item the top cover covered by a multi-layer insulation fabric (MLI). 
\end{itemize}

Each one of the 25 identical target modules (see figure \ref{fig:figmodule}) consist of 64 scintillation bars optically coupled to a multi-anode photo-multiplier (MAPM) with 64 separate readout channels. Each module is packaged in its own
1~mm thick Carbon Fiber Reinforced Polymer (CFRP) box which also contains the front-end boards with high
voltage (HV) distribution, pre-amplification and digitization. All
elements used in the construction are detailed in the following sections.

\subsection{Aluminum frame}
The aluminum frame is the main mechanical support of OBOX. The alloy AL 7075T7351 was chosen, among other reasons, for its low level of activatable
isotopes such as manganese.
As can be seen in figure \ref{fig:obox} it consists of four pockets surrounding a central grid. The four pockets, when viewed from the back of the instrument contain (in clock-wise order when looking towards the zenith direction): the central computer (CT),
the high voltage power supply (HV PS), the low voltage power supply (LV PS) and the cabling pocket with the connectors  for the cable harness to IBOX. The central grid serves as a mechanical support for the 25 detector modules (see Section~\ref{sec:module}). The aluminum frame also conducts the heat generated by modules and electronics to TG-2. Heat is furthermore lost through radiation through white paint on the exposed surfaces (see Section~\ref{sec:paint}.)
The Aluminum frame is suspended on top of TG-2 using shock dampers.
The shock dampers are constructed  connecting two coaxial cylinders within a hollow cylinder of rubber. This is a rather standard design that gives three-dimensional elastic connection between the central screw and the object to be suspended. The rubber material was provided by the Chinese space authorities, respecting their requirements for out-gassing and time degradation. 
A MLI curtain protects the rubber against direct solar irradiation.

\subsubsection{Venting}
Under atmospheric pressure, the aluminum frame and the modules contain several liters of air that must be vented
during the ascent. The pressure drop during a typical launch is rapid requiring the venting devices to be
able to release the internal pressure efficiently while at the same time not allowing external light to enter the instrument.
In total 4 venting devices, all designed and manufactured in Geneva university,
are present in OBOX: two venting devices on the aluminum grid and two on the carbon cover.
The two venting devices on the aluminum frame, located in quadrant 1 and 3, are formed by hollow screws.
The opening inside the screw has the shape of a maze which is painted black.
The hole size was calculated such that the speed of the escaping air will stay well below the speed of sound during ascent but, at the same time,
the size of the holes must be smaller then the Debye length of the plasma typical at our orbit (centimeters) to prevent ions to penetrate inside OBOX \citep{nasa}.
The two venting devices mounted on the carbon cover are simple hollow CFRP covers with a hole on the side.

\subsubsection{Paint}\label{sec:paint}
A special paint (GS121FD from the MAP French company) has been applied to the aluminum frame. The paint has a white color in visible light and  a grainy texture. Its emmisivity (0.8) and reflectivity (0.2) was chosen so that it radiates heat to the deep space. The Solar absorption coefficient
after application and drying is a modest 0.2. The cooling of OBOX is completely passive and relies mostly on the characteristic of this
paint. A secondary function of the applied paint is the protection of the aluminum frame against atomic oxygen chemical attack.

The painted assembly was simulated in a thermo-mechanical model and subjected to a thermo-vacuum test for verification. The test was performed on a properly painted model featuring the correct MLI covering and thus is representative of the thermal performance during the night part of the orbit. Initial thermal performance of the instrument in space is in agreement with calculations. The full thermal performance of POLAR in space will be addressed in a separate paper.

The paint used on the aluminum frame is fragile and can be very easily scratched. On FM, the paint was applied at the latest possible moment just before shipping to China. Last minute repairs of handling scratches were executed at the launch pad
just days before attaching OBOX to Tiangong~2.

\subsection{Top cover}
Between the target and the MLI, a cover provides light tightness and mechanical strength, and shields from low energy particles. It consists of CFRP of 1~mm thickness. Its main purpose is to passively reduce the flux of electrons with energies below 1~MeV which populate the radiation belts. Without shielding it was shown in \citep{Nicolas} that the high flux of these particles can induce random coincidences indistinguishable from Compton scattering photons.

The cover also provides mechanical stability to fix the 25 modules in place and reduce the vibrations encountered during launch.
The back end of each module is fixed to the cover by one M3 screw.

\subsection{MLI}
The unpainted CFRP surfaces of POLAR are covered by a multi-layer insulation foil (MLI).
The MLI is composed of 15 layers of 50 $\mu$m polyimide film. The top layer features a light reflection sheet
of 6 $\mu$m polyester. The goal of this MLI is to thermally isolate the target from its surrounding. In simulation, the MLI is considered as a perfect absorber for heat conduction in both directions. The MLI is also protecting again micrometeorite impacts and absorbs very low energy electrons.

\section{Target modules}\label{sec:module}
\begin{figure}
  \centering
  \includegraphics[width=\linewidth]{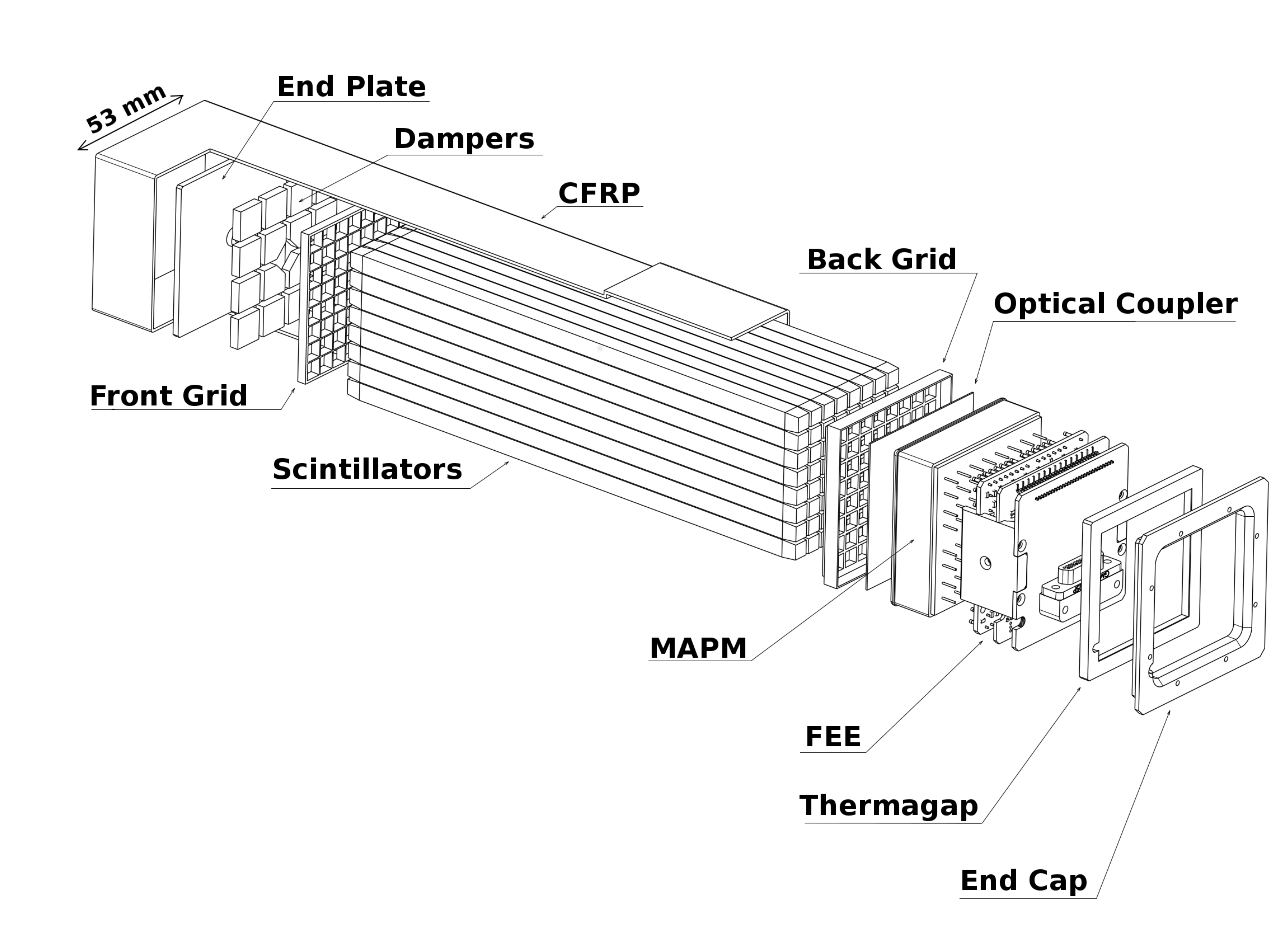}
  \caption{Exploding view of one module.}
  \label{fig:figmodule}
\end{figure}

For modularity and redundancy, the photon target is segmented in 25 modules. Each module (see Figure~\ref{fig:figmodule}.) is an independent sensitive unit
with its own plastic scintillator hodoscope and its own front-end electronic (FEE). The modules communicate only digitally with
the rest of the system. Each module is enclosed in its own container such that it forms a fully functional transportable object.

\subsection{Module shock absorbers}
The screw connecting the module to the CFRP cover connects to a plastic piece made of Polyether ether ketone (PEEK) (end plate, 2mm) placed at the top end inside  the module. Below the end plate is a bumper which was made of rubber material for the first QM which suffered catastrophic loss of all MAPM. Rubber is an almost uncompressible material with a Poisson modulus of 0.4999. Thus under longitudinal pressure,
rubber expands sideways but the same volume. If the rubber is laterally constraint it thus becomes in-compressible. This caused the failure of first QM to withstand shocks in the longitudinal direction.

The damping rubber (3mm) was thus cut into 16 square pieces with clearance between them. This allows for a side-way expansion of the rubber, efficiently dissipating the energy of lateral shocks.

This solution alone was calculated to be sufficient to cure the problems encountered in the shock test of QM. However a further safety margin was produced by several additional improvements. Rubber was replaced by Sorbothane (produced by Sorbothane, Incorporated) which has better dissipation properties. The rigidity of the aluminum grid holding modules was also increased by adding ridges.

Both the PEEK and the rubber absorbers are kept as thin as possible to minimize the absorption incoming photons. The materials add an extra 8\% attenuation for 50 keV photons coming from zenith.

\subsection{Baffles}
Two baffles (front baffle and MAPM baffle) made of Polyurethane (PU5286 from Huntsman Company) with 64 square holes are very precisely manufactured by additive technology. They mechanically fix the 64 bars in place. The front baffle connects to the shock absorbers described above and the bottom grid rests on the MAPM with the optical pad in-between. The grid structure of the baffles ensures that every scintillation bar is well centered on the structured cathode of the MAPM.

\subsection{Scintillating bars}
The active material in one module consists of 64 bars of scintillating material EJ-248M made by Eljen Technology, a scintillator type which is resistent to high temperatures.
The high temperature material was chosen as in the lab creeping of the non high temperature resistant version of the material at temperatures as low
as 40 degrees was observed and POLAR was required to be capable of storage temperatures up to 60 degrees.
The shape of the scintillators is a rectangular cuboid of length 176 mm and  $5.8\times5.8$ mm$^2$ surface dimensions. The attenuation length of visible light in the scintillation material is 1.6 meter and is not a significant concern for these short bars.

Both ends of each bar are truncated into a conical shape over the last 4~mm to end up with lateral dimension of 5$\times$5 mm${^2}$ which matches the size of the MAPM cathode pixel. The far end conical shape is useful for mechanical support and was shown (\cite{MerlinSim}) to enhance the
uniformity of the optical yield.

The scintillation yield of this material is one visible photon per 108~eV of electromagnetic energy deposition.
X-ray photons of interest for POLAR range between 10 keV and 500 keV in energy. These photons
interact with the bar material either by photo-electric effect or by Compton scattering.
In both cases an electron is produced in the scintillation material. The range of the produced electrons
is typically much lower than the dimension of the bar. Their complete energy is thus typically deposited in a single bar.
The light yield of scintillators is not fully linear but depends on the density of the energy deposition by Birks effect \citep{Birks}.
The POLAR Monte Carlo simulation takes into account this important detail \citep{MerlinSim}.

Scintillation photons are emitted isotropically and when reaching the bar surface under a shallow angle will be totally reflected. Such photons are transmitted to the end of the bar in an essentially loss-free way. 
The quality surface finish of the bars is essential to guarantee that the collection efficiency of the visible photon does not depend on the location of the energy deposition. The scintillation bars are produced in form of a plate with the correct 5.8 mm thickness. These plates were then diamond cut into bars. For two surfaces, the mirror finish is thus guaranteed by the production process. The two other surfaces were polished to ensure a mirror finish. The roughness of one bar was measured with a scanning micrometer device. The distribution of the normal angle was measured by differentiation of the height versus distance curve.
It was found that for the two good surfaces, the distribution of the normal angle is Gaussian with a width of 2.1 10$^{-5}$ rad.
For the manual polished surfaces a width of 2.6 10$^{-4}$ rad was found. 

Those measured values are used in the GEANT4 based Monte Carlo. The simulations have shown that the surface smoothness is an important parameter for the performance of POLAR. However, when the smoothness is better than 5 10$^{-4}$ rad results become insensitive to this parameter. The two conical ends of the bar are also machined and polished.

\subsection{Vikuiti reflective foil}
Photons leaving a scintillator bar, either laterally beyond the total reflection angle or longitudinally at the far end, must be prevented from entering into adjacent bars or getting lost. For this purpose each bars is wrapped with an opaque and highly reflective material.

As a high performance reflector the material ``Vikuiti Enhanced Specular Reflector Film (ESR)'' from 3M company was chosen \citep{3M}. Vikuiti provides mirror-like ultra-high reflectivity of greater than 98\% over the full visible spectrum and all incident angles. To not prevent total reflection at glancing angles, the Vikuiti is not attached to the bar.

\subsection{Optical coupling pad}
At the bottom end of the bar the optical photons should penetrate into the MAPM without reflection, using an optical coupling matching the indices of refraction of the materials.
Non-liquid soft materials with the correct index of refraction were investigated and a solution designed and successfully tested in space for the AMS detector \citep{AMS} was chosen. Soft transparent pads of 0.7 mm thickness were constructed by curing the two component material Dow Corning 93-500 \citep{Dow} at moderate temperature under vacuum.

The mechanics of the bar mounting ensures that the bars apply a small pressure to this interface material at all times even during vibrations. This ensures that there is no interface where internal reflection can take place.

Optical photons arriving at the MAPM end of the bars penetrating the interface material can diffuse sideways and hit neighboring anode pads. The probability depends on the exiting angle from the scintillator bars. This creates optical cross talk to neighboring channels, which is highly undesirable as it dilutes the signal, blurs the position accuracy and reduces the energy resolution. This optical cross talk is proportional to the total distance between the bar end and the cathode pad inside the MAPM. This distance is therefore minimized. It is furthermore important that the positioning baffle is opaque. This optical cross talk is inevitable and reduces the performance of the instrument.

\subsection{Multi-Anode Photo-Multiplier (MAPM)}
The MAPM transforms the optical photons from the scintillator into a measurable electrical charge. The multi-anode metal mesh photo-multiplier type HAMAMATSU R10551-00-M64 \citep{Hamamatsu} was chosen for POLAR. This photo-multiplier has 64 anode in a flat panel square assembly. It features 64 independent cathodes of dimensions 6.08$\times$6.08 mm, 12 dynodes and a grid and 64 independent anodes. Each channel can provide a gain of up to 1.5 $\cdot10^6$. With each MAPM, the HAMAMATSU company provides a measure of the cathode luminosity (typically 60 $\mu$A/lm), and a uniformity map. For POLAR a maximum gain ratio of 2 between all channels within a MAPM was specified. The best MAPM were selected based on the properties provided by HAMAMATSU for the POLAR flight model. The MAPM has an electrical cross talk at the percent level, negligible compared to the optical cross talk  mentioned above.

\subsection{Front-End Electronic (FEE)}
The front-end electronics (FEE) was designed with these particularities in mind (see also~\cite{Radek}):
\begin{itemize}
  \item It must be completely conduction cooled.
  \item It must work at atmospheric pressure as well as in vacuum. Operation at the most demanding
    pressure for high voltage sparking, around 1mbar, is not required.
  \item The lateral dimension must be compatible with the MAPM dimensions to enable 4 side buttability.
  \item The communication is fully digital.
  \item It can work and retain its performance (gain and dynamic range) in an extended temperature range (-40$^{\circ}$ to +60$^{\circ}$)
  \item It withstands 5 years of low Earth orbit (krad level) radiation without visible gain degradation.
  \item It can withstand random vibrations of the order of 20 $g$ and shocks of the order of 500 $g$.
\end{itemize}
The FEE consists of three stacked PCB boards held together from two sides by an aluminum flange that serves
as a mechanical structure and provides thermal contact to the rest of OBOX.
The three boards of the FEE are described below.

\subsubsection{HV distributor board}
The high voltage distribution board is situated directly behind the MAPM. All the MAPM pins connect directly to
the board by means of gold tulip connectors. The anode signal is capacitively decoupled with 330 pF and 100 $\Omega$ to ground.
The signal from dynode 12 is capacitively coupled with 4.7 nF and 10 k$\Omega$ to ground. 
The high voltage arrives to the board via a coaxial cable. The inner conductor of the coaxial cable is
directly soldered on the board. The shield of the coaxial cable is soldered but not
connected as it would make a ground loop with the local ground of the board (see Section \ref{sec:ground}).
The HV goes to the cathode through a 100 M$\Omega$ resistor. This line is capacitively coupled to ground by
three high voltage capacitors of 12 nF. All noise on this line couples directly to every anode output so this is a critical signal for performance.

The dynode high voltage divider consists of twelve 470~k$\Omega$ resistors. Further resistors and capacitors are used in the dynode chain for enhancing high rate performance.

The HV distributor board contains no active component but is densely populated with passive components and connections.
The power dissipation of is limited to 0.1 W, so that it can be easily conducted through the connectors and the aluminum flange.
The HV board is connected to the subsequent FPGA board through two SAMTEC 40 pin connectors.

\subsubsection{FPGA board}
The FPGA board takes care of the signal shaping, digitization, the trigger decision and the communication to the rest of the set-up.
The 64 anode signals arrive at a readout chip (VA) from the Viking family \citep{VA}. This chip is a 64 channel shaper, discriminator, sample and hold and analog multiplexer from the IDEAS company. The chip requires a -2.5~V, ground and -1.5~V power supply, which is provided by the third board of the FEE assembly described below. The chip die is directly glued to a gold contact on the PCB and connected by gold wire bonding.

The analog output of the Front-End chip is digitized by a 25~Ms/s 14~bit ADC. The sequencing of the readout, the trigger logic decisions and
the data compression are taken care of by a AGL250V2 FPGA (AGL) \citep{AGL}. The trigger is based on signals from very fast discriminators, which transmit to the AGL the multiplicity of channels over threshold. The FPGA receives within 30~ns a signal that says if there is at least 1 channel, 2 channels or more than 8 channels above threshold. The number of channels quoted here are the default values on a programmable DAC, these settings can be changed. All analog signals to the VA like the overall threshold are also programmable by DAC. The AGL implements the trigger logic and all the communication to the outside. 

As soon as one channel is above threshold, the 64 VA channels are put into hold. This state stays put until the higher level trigger takes a decision. If the decision is positive, a full readout occurs which lasts for 6~microseconds. if the decision is negative the VA chip resumes live operation. The AGL maintains also a counter to measure  dead time. A LS64 chip (produced by the same company which produced the VA) is used for level shifting of the 64 discriminator outputs between the non-matching voltage levels of the VA and the AGL. Connection to the LS64 is made by direct wire bonds. All wire-bounding schemes have been tested for vibration. 

The FPGA board also contains a digital thermometer, featuring a unique readable digital ID that is used to identify each FEE assembly.

\subsubsection{Power regulation and connector board}
The third  board of the FEE assembly provides a space qualified connector between the module and the rest of OBOX and to filter and recondition the voltage level received from the low voltage power supply.

\section{OBOX back-end electronics}
The low voltage power supply (LV PS) was specified to respect the desired size, operation, interface and space qualification requirements. These specifications were provided to ``Art of Technology'', a company based in Zurich \citep{AoT} for design, engineering and board production. The LV PS independently generates three voltages for each FEE (+3.3, +2 and -2.5). A single +28V input is provided to the LV PS from IBOX (see Section \ref{sec:IBOX}). The output voltages are stable to within 1\% and independently switchable.\\
The high voltage power supply board (HV PS) was produced by the same company that produced the LV PS.
The HV PS works in two stages. First a global high voltage level is produced (settable between 600~V and 1200~V),
subsequently 25 current limited sources (settable between 0 and 200 $\mu$A) are derived from this high voltage and provided to individual MAPMs. As the MAPM voltage supply is essentially an Ohmic load, a current source corresponds to a voltage source. 
The HV PS current sources makes heavy use of opto-coupler transistors to regulate the current. Initially worries existed that the transistor gain would change with irradiation, which would compromise the long term stability of the high voltage. A large batch was therefore irradiated up to 5 krad using the gamma-ray therapy facility of the Geneva hospital oncology department. This test have shown no important degradation.
The 25 FEE assemblies are connected to the HVPS by coaxial cables. They are connected to the CT and to the LV PS using five flexible PCB cables (flex).

\subsection{Central computer (CT)}\label{sec:CT}
\begin{figure}
  \centering
  \includegraphics[width=\linewidth]{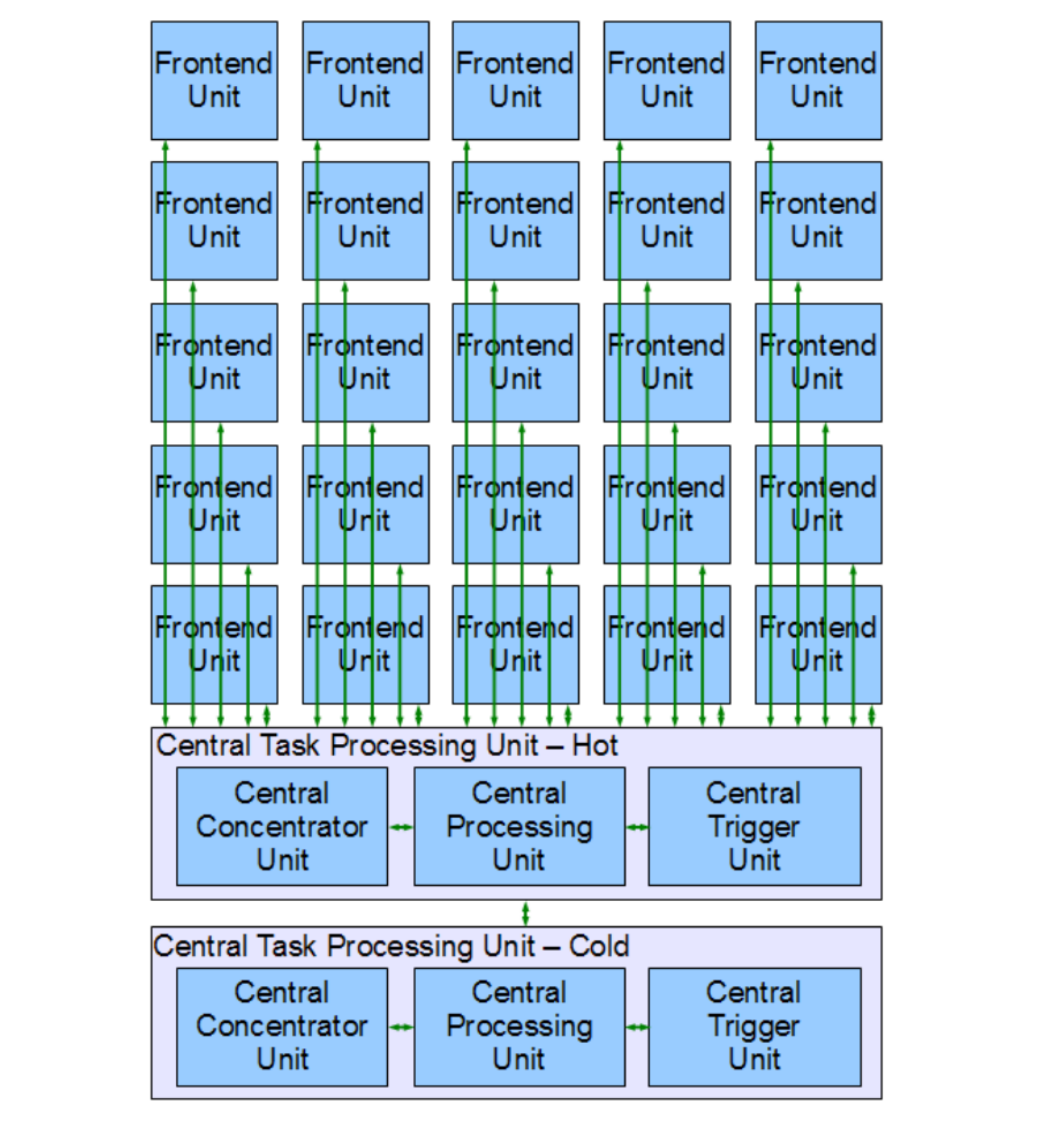}
  \caption{Functional block diagram of the POLAR CT.}
  \label{fig:CT}
\end{figure}

The central computer (CT) is described in more details in \cite{Dominik}.
The CT has several roles in POLAR:
\begin{itemize} 
\item It receives and treats all commands from IBOX.
\item It pilots the HV PS and the LV PS.
\item It synchronizes the FEEs by distributing a common clock.
\item It answers promptly to trigger request from all the 25 FEE and implements the overall event trigger logic. 
\item When events are in the FEE buffer these are read out and combined into science packets containing both the FEE events and the trigger decisions. 
\item Periodically all temperature sensors are read out as well as the status of the FEEs. The CT builds the housekeeping telemetry packets. 
\item It sends science and housekeeping telemetry packets to IBOX.
\end{itemize}
The CT tasks are distributed to three FPGAs: the computer FPGA (CPU), the trigger FPGA (TRIG) FPGA and the communication FPGA (COM). All FPGA are of the type A3PE3000L-FG484I. A block diagram of the CT functionality and connectivity is shown in Figure~\ref{fig:CT}.

\subsubsection{CT board}
The CT is constructed using surface mounted chips on a 8 layers PCB as described in \citep{Dominik}, providing two-fold redundancy. Apart from the three FPGAs it features a very stable
crystal clock, fifty Low Voltage Differential signaling (LVDS) level shifter to establish LVDS bidirectional communication with FEE assemblies,
one temperature sensor and SPI and I2C drivers to communicate with LV PS and HV PS. On power-up, the CT is in a in cold spare state.

\subsubsection{Connectors}
The CT receives or drives up to 500 real time signals from the FEE assemblies through the flex cables. For this purpose ten high density nano-size connectors are used (type 891-007-51S-A2-BST-1-J and 891-007-25S-A2-BST-1). These connectors are rather fragile and susceptible to short circuits between connections. The number of mating and unmating actions on the FM and FMS thus had to be strictly limited and documented and extreme care was taken for each connection to a flex cable. 

\subsubsection{CPU}
The CPU FPGA acts as a computer. It communicates through LVDS with IBOX, receives and interprets all commands from IBOX, is the master of the LV PS and HV PS, surveys all temperatures, builds housekeeping telemetry and sends prepared science packets to IBOX. The CT is also able to work completely independently when booted.

\subsubsection{Trigger}
The trigger FPGA implements the instrument trigger logic. As soon as a FEE has at least one channel with a signal over threshold, the FEE sends a demand for treatment to the trigger FPGA. Subsequently the FPGA waits 100~ns to see if a demand comes from another module. It then sends a positive or a negative answer to all waiting modules. The answer is based on the multiplicity communicated by each waiting FEE. If one waiting FEE has a large multiplicity, the event is rejected apart from a pre-selected fraction. The pre-scaling value can be set by the user. 

If at least two FEEs have a multiplicity higher than one but below the high multiplicity threshold, the event is accepted. If at least one FEE has a multiplicity of two, the event is also accepted. If only one module is waiting and it has a multiplicity of 1 the event is pre-scaled. Also for this the fraction of stored events can be set by the user. 

The trigger FPGA distributes a master clock to every FEE. The clock signal is used by the FEE to time-stamp the events and to compute a dead time. Each time the trigger FPGA takes a decision it creates a time-stamped trigger packet explaining the decision. In standard operation the trigger FPGA does not produce trigger packets for rejected events. The trigger FPGA also measures its own dead time. However the dead time in POLAR is dominated by the dead time from the FEEs and the trigger FPGA dead time is negligible.

\subsubsection{Communication}
The communication FPGA continuously checks the 25 FEEs for events waiting to be processed. If this is the case, the event is read out and put into a FIFO for the trigger FPGA to decide and forward it to IBOX. The communication FPGA can sustain 25 communications in parallel. If the FIFO is full, the trigger FPGA will not accept further events.

\subsubsection{Science data}
The CT is just moving packets from FEE to IBOX communication. However, it adds to this data stream a data stream coming from the trigger decision one
packet per positive event trigger decision. A physical event corresponds to one trigger packet together with 1 to 25 module packets.

\subsubsection{Clock distribution}
The CT has also the task of maintaining an absolute time reference. The CT features a very stable ($10^{-6}$) quartz oscillator. Oscillations are counted
at a 12.5~MHz frequency and this count is attached to every event. IBOX is sending every two minutes to OBOX a synchronization (SYNC) command that is
very precisely timed versus the arrival
of a pulse per second pulse from a Global Positioning system (GPS) receiver. OBOX is storing the counter value each time a SYNC command arrives and this value is sent in the housekeeping telemetry.

\subsubsection{Commanding}
The CT receives commands from IBOX, decodes them and implements them. If the command is decoded and understood correctly an acknowledge packet
is created for every received command. Most of the commands just write a parameter to a register which will be used for FEE initialization.
The CT has four different states: Booting, Data Acquisition (DAQ), maintenance, calibration.
The commands ``go the maintenance mode'', ``SYNC'' and ``read register'' can be received at any time. All other commands (``write register'', ``start LV PS'',
``start HV PS'', ``go to DAQ mode'', ``go to calibration mode'' can be received only in maintenance mode.

\subsection{Heaters}
Heaters in the form of a resistance printed on a kapton foil are glued on the inside of the four pockets housing the OBOX back-end electronics in the aluminum frame. The goal is to provide heating power inside OBOX in case of low temperatures which could damage the instrument. Each heater resistance is 160 $\Omega$ for a total resistance of 40~$\Omega$ delivering 20W from the nominal 28V supply. The heaters are piloted from IBOX. IBOX is also measuring the temperature of the aluminum frame using two PT100 resistors and automatically supplies power when the temperatures drop below a settable threshold.

\subsubsection{Outside Connectors}\label{sec:connectors}
OBOX features 3 connectors on the outside. One (X2) is the power connector through which OBOX receives the main 28V hot and 28V cold power from IBOX. Each electrical
connections uses two conductors for redundancy. The hot and cold power supply are never used at the same time. The X2 connector brings also power to the internal heaters and connections to the two PT100 temperature measuring resistances.

The X1 connector provides four bidirectional LVDS communication channels to the platform. Two bidirectional channels are connected to the hot CT and two to the cold CT. A third connector is used to flash the 3 FPGAs of the hot and the 3 FPGAs of the cold CT. On FM, this connector was masked just before the launch because in flight condition it does not connect to anything.

\subsubsection{Filters}
During EMC tests some filtering turned out to be necessary on the main 28V power supply to withstand the rather drastic transients that they have to cope with. All power lines go first through a ferrit coil, then all 28V power lines go through a filter of type Tusonix 4209-053LF. The 28V lines are capacitively coupled to ground through a four-capacitor circuit with 2 parallel of 2 serial capacitors, which can withstand any single capacitor failure and some double capacitor failures. The equivalent capacitance is 2.2 $\mu$F rated up to 100V.

\section{Redundancy Scheme}
The instrument was designed to be fully redundant to allow for small malfunctions, by eliminating single point failures as far as possible. The split into 25 identical target modules (see Section \ref{sec:module}) already provides redundancy. In case one module stops working, POLAR loses between
4\% (The surface of one out of 25 modules is lost) and 8\% (this module cannot participate to 2 modules events) of its sensitivity depending on position.

The LVPS is twofold redundant in its first stage in addition to the 25-fold redundancy of the target modules. The first stage of the HVPS supply has a constant hot spare and the 25 current limiters are duplicated as cold spares.

There are two identical CT (see Section \ref{sec:CT}) in cold redundancy. All the power cabling is doubled. Communication between OBOX and IBOX also has a cold spare. Both the cold and the hot communication channels are implemented with two redundant channels. The first channel which passes a validated packet is retained as the active channel.

Most of the electronic components used are Commercial off-the-shelf (COTS) but industrial quality is used when existing.
Some components are determined to be suceptible to create a major failure, for those military level components are used
mostly in the HV power supply. Components
like the VA and the LS64 where qualified by the collaboration.

\section{Electrical and thermal coupling}
The coupling between the module and the aluminum frame must provide both a good ground connection and good thermal coupling.
It must furthermore ensure that the modules stay aligned even after the vibration of launch. For these purposes the material Therm-a-gap  \citep{Parker} was chosen, which remains soft under compression and has a high thermal and electrical conductivity. The thickness of the Therm-a-gap is 4~mm and pressure is applied using the screw connecting the module to the carbon cover, controlled by the torque applied to the screw.

\subsection{Grounding scheme}\label{sec:ground}
The signals to be measured by POLAR electronics are of the order of fC, such that an optimum grounding scheme is required. 
While in-orbit OBOX is exposed to free electrons and atomic oxygen.
This means that high resistivity material should be protected and every conductor should be ohmically connected to ground.

The reference ground is the ground from the power cable. The aluminum grid is connected to ground at a single point with a 1k$\Omega$ resistor. The CFRP OBOX cover, which is an electric conductor, is connected to the aluminum frame. This creates a
full Faraday cage around OBOX.

All communication uses LVDS signals. This way the communication relies only on the difference between two voltages
and not on a level versus ground.

\section{Calibration sources}
\begin{figure}
  \centering
  \includegraphics[scale=0.5]{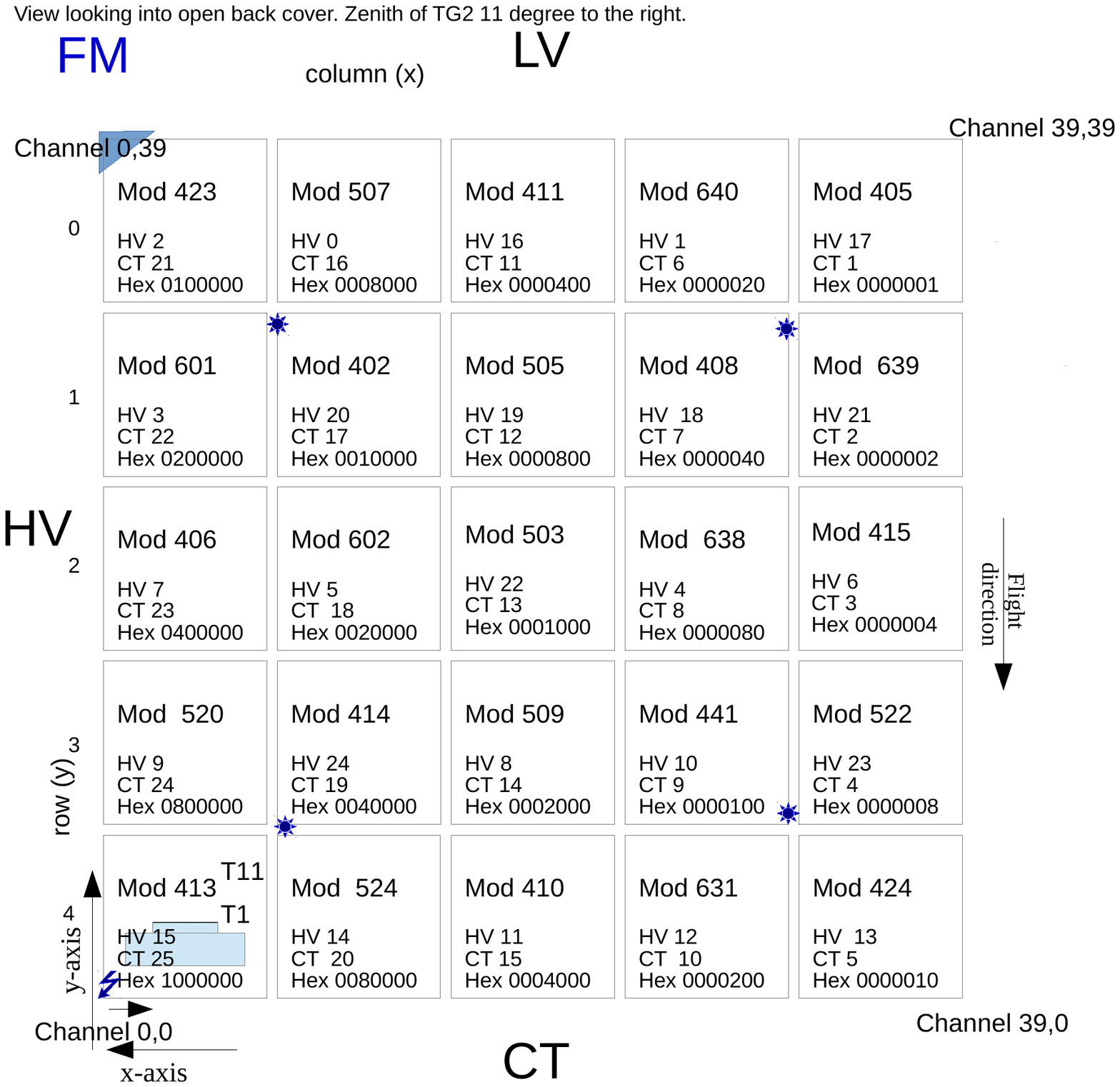}
  \caption{Numbering of target modules in the Flight Model. The z axis, corresponding to the zenith direction under normal flight conditions, points into the page.}
  \label{fig:back_module}
\end{figure}

Weak radioactive sources of $^{22}$Na where prepared by cutting a 1~MBq $^{22}$Na mother source. The mother source was ground to dust, which was recovered using adhesive kapton tape.
Each of the sources was glued inside an L-shaped piece of copper of 200~$\mu$m thickness. The copper prevents any positron from escaping and provides a pure source of 511~keV photon pairs. 
Eight sources with appropriate activities between 100 and 200~Bq were selected and glued in FMS and FM
into positions indicated by the star symbols in Figure~\ref{fig:back_module}.

$^{22}$Na was chosen because it creates two coincident back to back photons of 511 keV. This allows for the event topology of
two energy depositions lying on a line passing near the calibration source
to be used to extract those rare events.

\section{IBOX}\label{sec:IBOX}
\begin{figure}
  \centering
  \includegraphics[width=\linewidth]{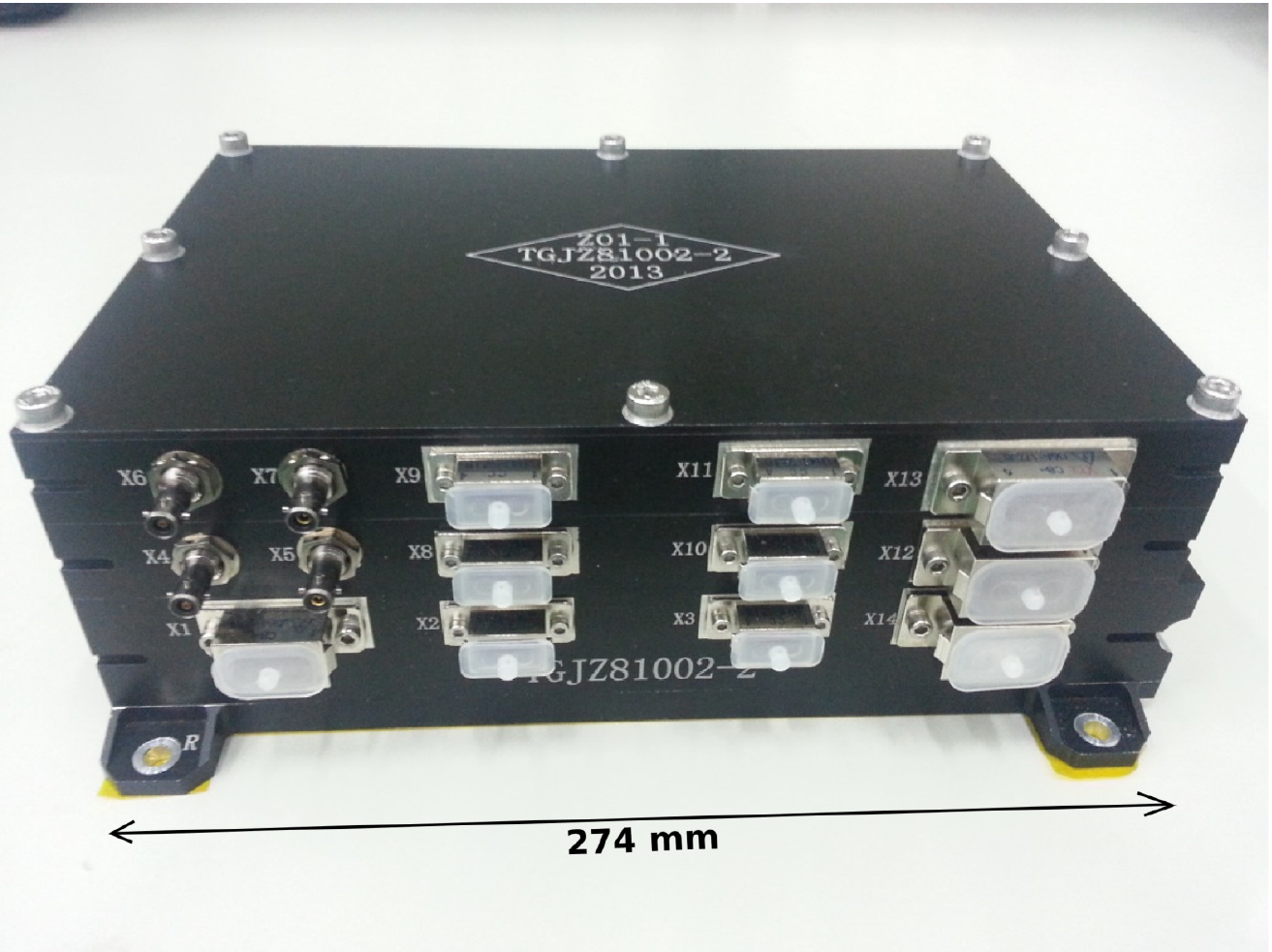}
  \caption{Image of IBOX Flight Model.}
  \label{fig:figibox}
\end{figure}

IBOX is mounted in the interior of TG-2, the flight model is shown in Figure~\ref{fig:figibox}. The two main parts of IBOX are a control unit and a low-voltage power supply. The control unit is responsible for the data management of POLAR
using LVDS bus communication with the satellite platform and OBOX.
Slow control is monitored using a MIL-STD-1553b (1553b)
communication channel.
It furthermore handles reception and execution of commands uploaded to IBOX from ground and the overall system monitoring and handling.
The low-voltage power supply provides all required voltages for both OBOX and IBOX itself. IBOX is assembled with a food steamer structure, of  5 layers in total.  From top to bottom: the top cover, a hot control unit, a cold control unit, the low-voltage power supply and a floor plate. The gross weight is 3.7~kg and the size is $247 \times 160 \times85$~mm$^3$.\\

IBOX contains two software configuration items: microcomputer software and FPGA firmware. The microcontroler software is mainly responsible for the communication of the 1553b bus,
undertaking the data interaction between the satellite platform and FPGA logic. The FPGA firmware, as the control core of IBOX, receives housekeeping data and science data from OBOX, offers relevant telemetry information to the satellite platform, manages all command injections, gives signals of timing calibration, and judges entrance and exits from the South Atlantic Anomaly (SAA) area where the HV of OBOX is switched on or off. The SAA
region is defined as a programmable rectangle in longitude latitude. During operation, this was seen to be too restrictive and
ground commands were used to turn on and off HV to create a polygonal SAA region (see also chapter on performances).\\
In the thermal design of IBOX heat is dissipated through conduction through the bottom plate which is in thermal contact with the satellite platform. The bottom plate has undergone oxidation treatment for electrical conductivity. The other surfaces are blackened to improve the energy contact with the platform through radiation while improving corrosion resistance of the equipment. As the power consumption of IBOX is mainly dissipated through contact conduction the low-voltage module is placed at the lowest layer, and all heating devices are directly attached to the bottom plate and the side walls. A thermally conductive silicone grease is applied between them.

\section{Thermal design}
OBOX is passively cooled using the white paint on the aluminum frame and the contact through the shock dampers
to the Tiangong~2 platform. In the design the Tiangong~2 platform was considered as an infinite heat bath of constant temperature. 
All the non-painted parts of POLAR are shielded with MLI and therefore don't participate in the thermal balance. When running normal data acquisition, OBOX dissipates about 45~W.
If needed and if there is enough power available from the station, heaters can be switched on to provide 20~W to the internal side of the aluminum frame pocket.
The thermal concept was fully simulated using ABAQUS 6.10. To calibrate the simulations, some of the thermal conduction paths were measured using mock-up models and with the first test modules.

Each FEE dissipates 1~W. The power in a module is essentially dissipated by the VA and the FPGA. To insure good thermal path, the VA is glued to a gold pad on the PCB.
This gold pad is connected through a large number of vias to an internal copper plane in the middle of the PCB. This plane occupies the full surface of the PCB. This central copper plane is in thermal contact with the aluminum flange of the FEE which brings the heat to the top of the FEE. It was found that the thermal transport can be enhanced by  equipping the two remaining sides of the PCB with a pyrolitic carbon foil (SIGMA\_ALDRICH MERCK). Those foils decrease the temperature gradient between the VA PCB and the top of the FEE by several degrees.

The LV PS efficiency is overall 80\% thanks to the usage of switching power supplies. The 10W generated here are easily dissipated by the large PCB-to-aluminum contact that the LV PS holder provides. A small dependency of the total power consumption with temperature can be attributed to a degradation of the efficiency with temperature. The HV PS additionally dissipates around 10~W and again has ample contact between PCB holder and the aluminum frame. The CT typically dissipates 5~W depending if it is idle or running the full data acquisition. Each FPGA dissipates about 1.5~W which are evacuated by the contact to the PCB. This PCB has a full layer devoted to thermal conduction. The rest of the power is dissipated by the LVDS converters.

All the heat that is not dissipated through the paint is conducted to the rim of the aluminum frame. The connection of the aluminum to
Tiangong~2 using the shock dampers is not very efficient in terms of heat transfer. Four flexible cooling plates made of copper enhance the thermal connection and don't interfere with the vibration damping.

\section{Basic performance parameters during first orbits of operation}
\begin{figure}
  \centering
  \includegraphics[width=\linewidth]{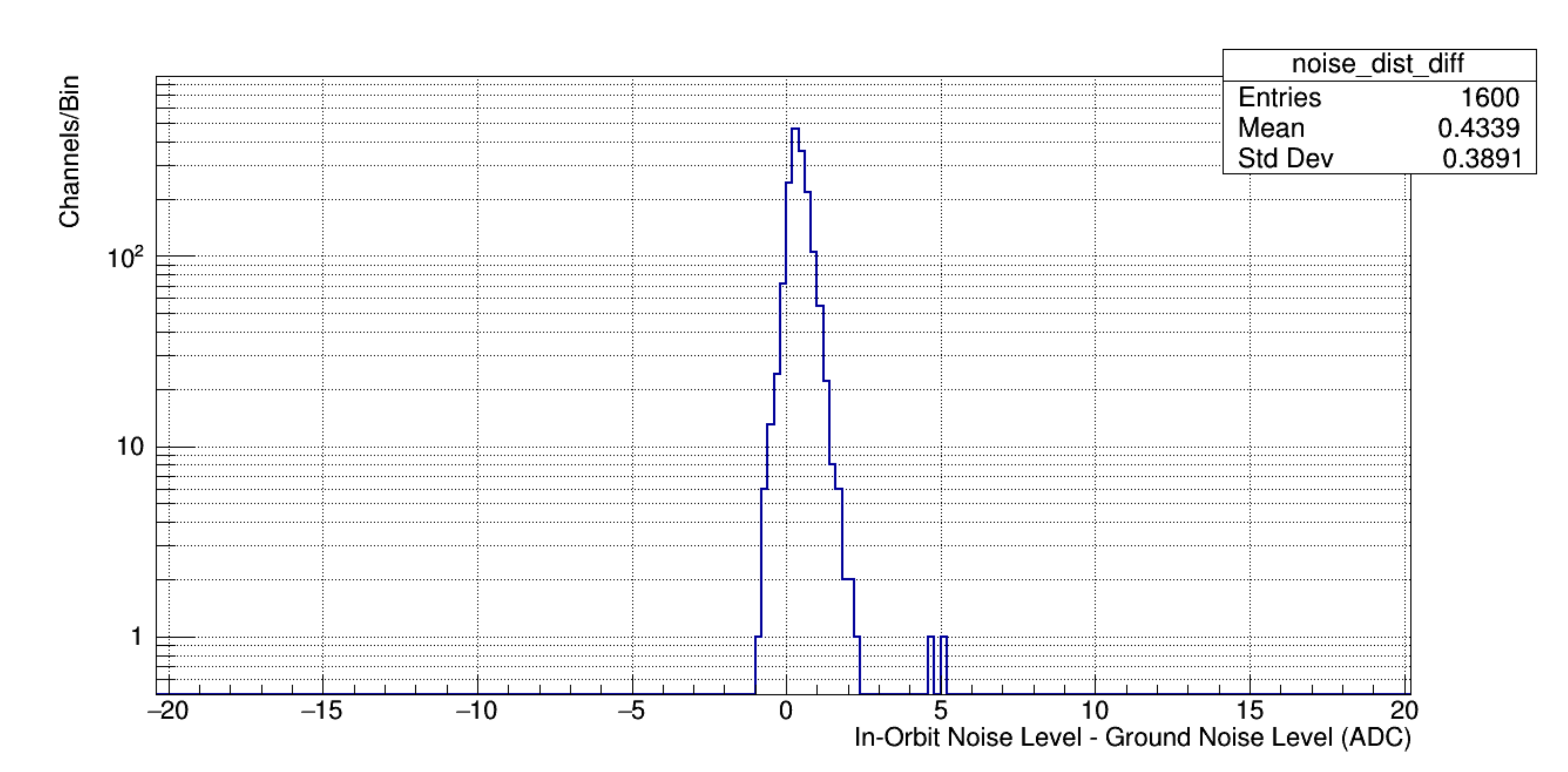}
  \caption{Intrinsic noise difference (difference between in orbit and at the launch pad) distribution. In orbit our intrinsic noise is 0.4 ADC higher then on the pad.}
  \label{fig:fignoise}
\end{figure}

\begin{figure}
  \centering
  \includegraphics[width=\linewidth]{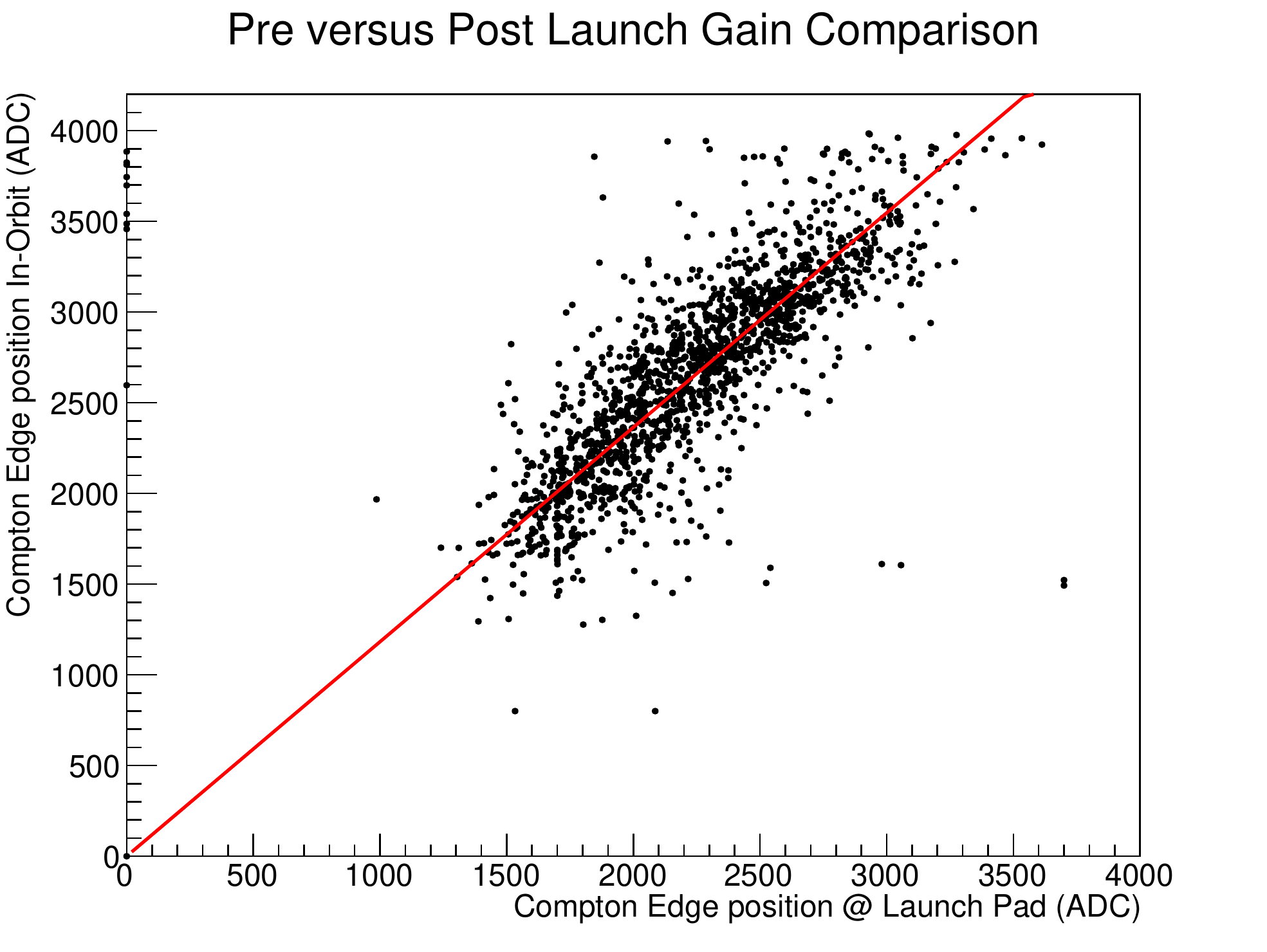}
  \caption{ADC location of the $^{22}$Na Compton Edge before and after
    the launch for the same High Voltage.
    The fitted line is forced to pass
    through the origin. It has a slope of 1.18. The slope is different than 1 because
    the data on Earth was taken at a temperature twenty degree higher than in space, compatible with the
    measured mean of -1\% gain change by degree.}
  \label{fig:figce}
\end{figure}

\begin{figure}
  \centering
  \includegraphics[width=\linewidth]{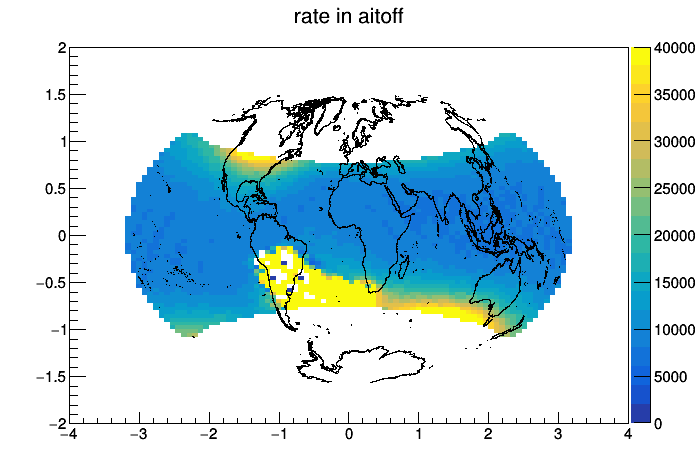}
  \caption{Rate of raw event trigger in Hz}
  \label{fig:figrate}
\end{figure}

\begin{figure}
  \centering
  \includegraphics[width=\linewidth]{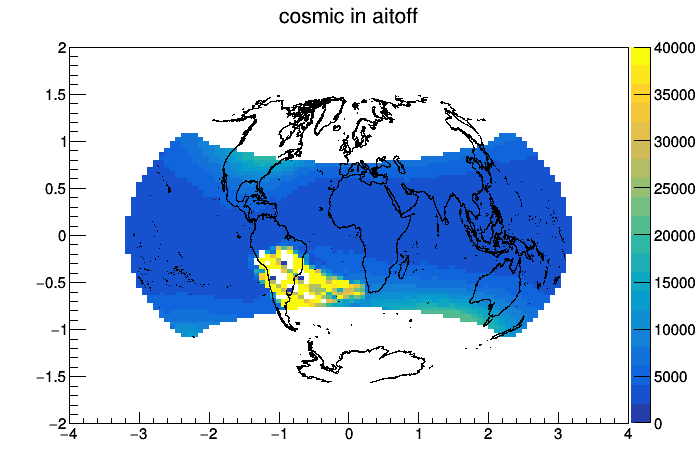}
  \caption{Cosmic raw trigger rate in Hz}
  \label{fig:figcosmic}
\end{figure}

The pre-orbit performances compared with the Monte Carlo were addressed in another paper \cite{MerlinSim}.
The full in-orbit performances and operation will be addressed in a future paper \cite{Yongjie}.
We display here some essential parameters of the detector that show that it performs as expected in space.\\
Figure~\ref{fig:fignoise} shows that the intrinsic noise of the instrument did not change
dramatically after the launch. As explained in \cite{MerlinSim}, the noise of the instrument consists of two parts:
a correlated noise (shared by all bars of the same module) and an intrinsic noise. The correlated
noise is about three time larger than the intrinsic noise but it can be measured event by event and subtracted.
A couple of bars became noisier and could be excluded from analysis without significantly affecting the instrument performance.\\
Figure~\ref{fig:figce} shows that the gain of the instrument did not change after the launch. The correlation is not 1 to 1 because the temperature
in orbit is not identical to the one on the launch pad. Some outliers are due to Compton edge fitting not converging because of the low
statistics collected on the launch pad.\\
Figure~\ref{fig:figrate} and \ref{fig:figcosmic} show the mean rate of raw events and cosmic ray triggers as a function of the longitude and latitude.
Rate of events is the rate of any event that passes at least a single bar threshold. Cosmic rays trigger are events with more
then 8~bars passing the threshold.

\begin{figure}
  \centering
  \includegraphics[width=\linewidth]{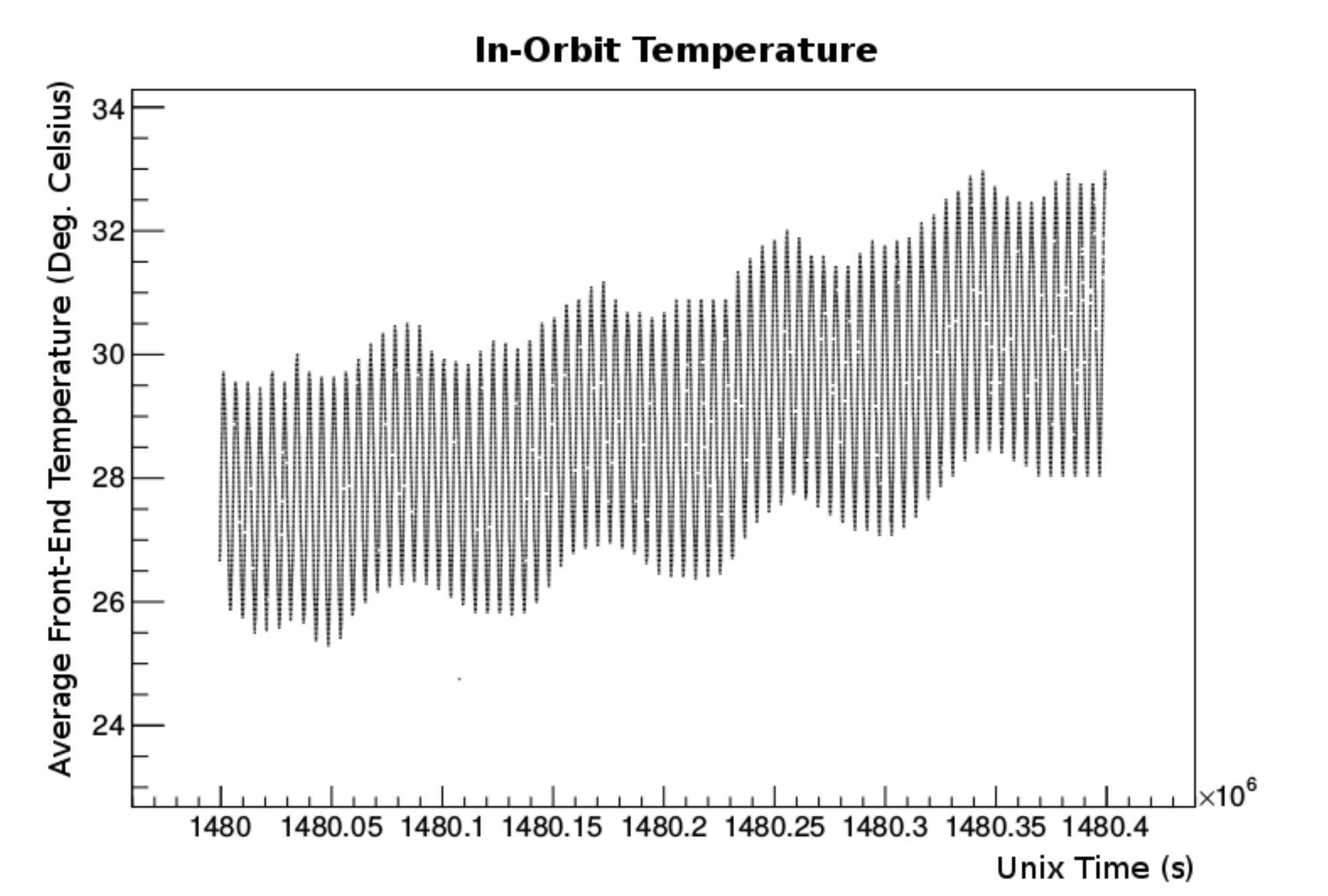}
  \caption{Mean temperature of all modules versus UNIX time.}
  \label{fig:figtemp}
\end{figure}

The SAA and the polar cap region are clearly visible. The SAA induces a dead time of
the instrument of less then 10\%.
Due to failure of commanding during a couple of orbits,
we do have some rate measurements inside the SAA.\\
Figure~\ref{fig:figtemp} shows five random days of temperature measurements. What is shown is the
mean temperature of all modules versus time. Temperature is driven by the solar irradiation that depends on the orbit and on the attitude of the spacecraft.
The fast temperature variation is due to the 90 minutes orbit.
The slow variation is due to attitude and season changes. The measured time profile correspond well with our thermal simulations.

\section{Conclusions}
The flight model and the flight spare were sent to China and accepted for flight during Summer 2016. The flight model was installed on the Tiangong~2 space laboratory in August 2016. POLAR together with Tiangong~2 where launched into space by a Long March 2F rocket on September 15 2016 at 14:04 UTC.

POLAR was successfully switched on and gave immediately satisfactory results. The thermal behavior was as expected.
Using the heaters, the modules of OBOX can be maintained within 5$^\circ$ around a mean temperature of 30$^\circ$. 

At the time of writing POLAR has already seen several tens of GRB and some of them are suitable for a polarization measurement. In-flight performance and scientific results will be published elsewhere.
\section{Funding}
Funding:This work was supported by the National Natural Science Foundation
of China [grant number 11503028]; Swiss National Science Foundation;
Swiss Space Office [ESA PRODEX program]; National Science Center Poland [grants 2015/17/N/ST9/03556].

\end{document}